\documentclass[12pt]{iopart}
\usepackage{epsfig,iopams}

\newcommand{\sub}[1]{_{\mbox{\scriptsize{#1}}}}  
\newcommand{\uv}[1]{\hat{\mathbf{#1}}} 

\begin{document}

\title{Optical excitation of nonlinear spin waves}
\author{J.~M. McGuirk and L.~F. Zajiczek}
\address{Department of Physics \\
Simon Fraser University, Burnaby, BC V5A 1S6, Canada}
\ead{jmcguirk@sfu.ca}
\date{\today}

\begin{abstract}
We demonstrate a technique for exciting spin waves in an ultracold gas of $^{87}$Rb atoms based on tunable AC Stark potentials.  This technique allows us to excite normal modes of spin waves with arbitrary amplitudes in the trapped gas, including dipole, quadrupole, octupole, and hexadecapole modes.  These modes exhibit strong nonlinearities, which manifest as amplitude dependence of the excitation frequencies and departure from sinusoidal behavior.  Our results are in good agreement with a full treatment of a quantum Boltzmann transport equation.
\end{abstract}


\section{Introduction}

Spin waves in ultracold atomic gases offer a new regime for exploring quantum dynamics.  Strikingly, applying a brief coupling between two internal quantum states in an otherwise classical gas can excite macroscopic collective quantum dynamics that persist for orders of magnitude longer than the initial perturbation.  Furthermore, many experiments must go to great lengths to achieve nonlinear couplings for studying more complicated behaviors than those present in simpler linear systems, whereas in the ultracold gas system, it is nearly impossible \emph{not} to drive the system into a nonlinear regime.

The first observations of spin waves in a fluid were in spin-polarized H and later in liquid $^3$He and $^3$He-$^4$He mixtures \cite{firstwaves}.  The high density and strong coupling between particles meant that these systems were deep in the hydrodynamic regime of the spin wave spectrum.  Thus, they gave rise only to linear excitations, and the spin orientations never diverged far from equilibrium.  The discovery of spin waves in ultracold gases opened the door to more detailed studies of these excitations due to the low density, ease of manipulation, and imaging possibilities of such gases \cite{lewando2002,mcguirk2002}.  It was immediately obvious that spin waves in ultracold gases were almost intrinsically nonlinear, since they occurred between the hydrodynamic and collisionless limits.  They showed large deviations from spin equilibrium, the spin states almost fully separating spatially, and strong amplitude dependence on the spin wave frequencies.  These nonlinearities complicated analyses, as simplifying linear approximations were not appropriate.  Much care was taken to eliminate the effect of nonlinearities by extrapolating to low amplitude limits, but the nonlinear behavior itself was not studied.

In this work, we specifically study the nonlinear nature of spin waves in an ultracold trapped alkali gas.  We do so by controllably exciting different spin wave modes with arbitrary amplitude and studying the frequency shifts produced.  Previous work on spin waves in trapped atomic gases relied on parameters intrinsic to the trapped atom system to excite spin waves -- in particular density-dependent mean-field shifts as well as the optical or magnetic trapping potential itself \cite{mcguirk2002,thomas2008,deutsch2010} .  Here we eliminate this driving mechanism and instead add an external optical potential relying on the AC Stark shift, so that spin wave amplitudes are decoupled from the trap parameters and arbitrary modes can be excited.

With this new method, we not only excite the quadrupole mode (mode number $d=2$) observed in \cite{mcguirk2002}, but we also demonstrate the flexibility of optical excitation of spin waves by driving dipole, octupole, and hexadecapole modes ($d=1,3,$ and $4$ respectively), which have not been previously observed in ultracold gas systems. We excite these modes with varying amplitudes and show that the excitation frequencies depend strongly on the amplitude, so much so that in some instances it is virtually impossible to excite spin waves in the linear regime where there is no amplitude dependence.  This behavior is in good agreement with numerical simulations of a Boltzmann transport equation.  Finally, we highlight the nonlinearity of this system by demonstrating that even after short time periods after beginning the excitation, during which the spin perturbations have not yet become large, the spin dynamics are inconsistent with those predicted by a simplified linear analysis.

\section{Spin wave theory}

Spin waves are macroscopic, collective excitations of atomic spin vectors that arise from the identical spin rotation effect (ISRE) due to exchange scattering between indistinguishable particles in an inhomogeneous potential \cite{oldtheory}.  The spin system we consider is a harmonically confined ultracold gas of spin-1/2 atoms, with spin states $|1\rangle$  and $|2\rangle$ coupled electromagnetically.  Atoms in an equal coherent superposition of the two spin states, $|\psi\rangle = (|1\rangle + \rme^{i\phi} |2\rangle)/\sqrt{2}$, in a spatially varying differential potential $U\sub{diff} (\bi{r})$ will acquire a spatially dependent phase, $\phi(\bi{r},t) = U\sub{diff} (\bi{r}) t/\hbar$.  This phase represents a transverse rotation of the local spin vector $\bi{S} (\bi{r})$.  As an atom oscillates in the confining potential, it undergoes coherent exchange scattering with other atoms having slightly different spin vectors.  Indistinguishability of quantum particles requires the scattered wavefunction to be symmetrized with respect to forward (unscattered) and backwards scattered terms, which leads to a rotation of the spin vectors of the two interacting particles about their combined total spin.  This ISRE has been observed in both bosonic and fermionic systems \cite{lewando2002,thomas2008}.  The ISRE gives rise to coherent spin currents, and the atomic spins undergo spatiotemporal oscillations of the spin vectors, \emph{i.e.} spin waves, about their equilibrium spin configuration as determined by $U\sub{diff}$.

Spin dynamics in this system are described by a Boltzmann spin transport equation \cite{levitov2002,fuchs2002,williams2002}.  Since the geometry of our atom trap is highly elongated in one direction ($z$), we consider only the one-dimensional evolution of the spin position and momentum distribution $\bsigma(p,z,t)$ and the experimentally accessible total spatial spin distribution $\bi{S}(z,t) = \int\,dp\;\bsigma(p,z,t)/2\pi \hbar$.  The transverse components of $\bi{S}$, $S_u$ and $S_v$, represent the spin coherence or the superposition phase $\phi$, and the longitudinal spin component represents the spin differential between states $|1\rangle$ and  $|2\rangle$, $S_w = N_1 - N_2$.  We adopt the notation of \cite{nikuni2003} and write the quantum Boltzmann equation as
\begin{equation} \label{boltzmann}
\frac{\partial \bsigma}{\partial t} + \frac{p}{m}\frac{\partial \bsigma}{\partial z} - \frac{\partial U\sub{ext}}{\partial z} \frac{\partial \bsigma}{\partial p} -
    \bOmega \times \bsigma = \frac{\partial \bsigma}{\partial t}{\Bigg |}\sub{1D},
\end{equation}
where we have suppressed the explicit dependence on $(p,z,t)$ for clarity.  The first three terms are the total time derivative of $\bsigma$, $U\sub{ext}$ is the harmonic magnetic trapping potential, and the right hand term represents a radially averaged 1-D collision term proportional to the elastic scattering rate.  We employ the analytical relaxation time approximation for the collision integral derived in \cite{nikuni2003}.  The remaining term is a spin torque that couples the components of the spin vector via the differential potential and mean field coupling, and is given by
 \begin{equation}
\bOmega = ( U\sub{diff}\uv{w} + g\bi{S} ) /\hbar,
\end{equation}
with $g = 4\pi\hbar^2 a/m$ for mass $m$ and s-wave scattering length $a$.\footnote{The scattering lengths $a_{ij}$ between states $|i\rangle$ and $|j\rangle$ ($i,j = 1$ or 2) vary by only a few percent in $^{87}$Rb.  This difference has only a minor effect on the dynamics and is not important for effects described herein, in contrast to phase separation in Bose-Einstein condensates, which occurs to minimize mean-field energy.}  It is this term that drives the spin rotations and provides the nonlinearity.

To make the physics behind these spin oscillations more accessible, \cite{nikuni2003} performed a small amplitude moment expansion of the spin distribution, \cite{levitov2002} considered the excitations in the hydrodynamic limit where spin perturbations are never far from equilibrium values, and \cite{mullin2005} compared the two methods.  These linear approximations give analytical results that allow for a better conceptual understanding of the phenomenon, and they accurately predict the small amplitude behavior of the spin excitations.  However, linear approximations fail to capture the full dynamics of large amplitude spin waves, and equation~(\ref{boltzmann}) must be treated numerically as was done in \cite{fuchs2002,williams2002}.  The observed nonlinear spin wave behaviors that are predicted by numerical solutions to equation~(\ref{boltzmann}) include amplitude dependent frequencies, frequency chirps as the excitations damp to equilibrium, and nonsinusoidal initial growth of $S_w$ at the onset of the excitations.

\section{Spin wave apparatus}

The system in which we study nonlinear spin waves consists of two hyperfine ground states of $^{87}$Rb, which comprise a pseudo-spin 1/2 doublet.  The two states,  $|1\rangle = |F=1, m_F=-1\rangle$ and $|2\rangle = |F=2, m_F=1\rangle$, are coupled via a two-photon microwave transition using a 6.8~GHz microwave photon and a 3.14~MHz radiofrequency (rf) photon, both detuned 700~kHz from the intermediate $|F=2, m_F=0\rangle$ state. The two-photon Rabi frequency is 3.2~kHz.  The atoms are confined in a hybrid Ioffe-Pritchard magnetic trap similar to that described in \cite{lewando2003a}.  The trap is cylindrically symmetric and elongated in the axial ($z$) direction with an aspect ratio of 37:1 (axial and radial frequencies 6.7~Hz and 247~Hz respectively), which allows us to radially average the spin profiles and treat the dynamics as one-dimensional.  We use rf evaporative cooling to prepare a sample of atoms at temperature $T=650$~nK and peak density $n_0 = 2.5 \times 10^{19}$~m$^{-3}$.  These conditions are $\sim1.5$ times the critical temperature for Bose-Einstein condensation, and the external degrees of freedom are well described as an ideal gas if one considers the gas as a single species and ignores the spin degree of freedom.  The spin distribution for all experiments described herein is initialized by placing the atoms into a coherent superposition of  $|1\rangle$ and  $|2\rangle$ via a $\pi/2$ pulse, corresponding to a spatially uniform spin state $\bi{S}(z) = n(z) \uv{u}$.

The differential potential experienced by the atoms has three contributions, as noted above: the magnetic confining potential, the mean field shift, and an additional optical potential.  By tuning the magnetic field, the differential mean field shift can be almost exactly canceled by the Zeeman shift to give a flat relative potential (see \cite{lewando2002}).  In the absence of any other inhomogeneities, no spin waves will arise in this configuration (although spin waves can still propagate in a uniform potential).

\begin{figure}
\begin{centering}
\leavevmode \epsfxsize=2.6in
\epsffile{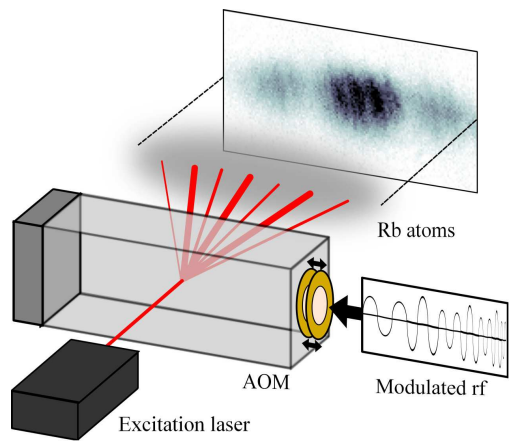}
\caption{\label{fig:apparatus} Schematic of the apparatus used for optical excitation of spin waves (not to scale), including an absorption image of the $|1\rangle$ state near the peak of a hexadecapole oscillation featuring three local maxima ($\bi{S}$ rotated towards $|1\rangle$) and two local minima ($\bi{S}$ rotated towards $|2\rangle$).}
\end{centering}
\end{figure}

We add a focused, spatially modulated laser to create tunable differential optical potentials and excite arbitrary spin wave modes (figure~\ref{fig:apparatus}).  The AC Stark shift for a far detuned laser creates an optical dipole potential $U\sub{dip} \sim I(\bi{r})/\Delta $, for laser intensity $I(\bi{r})$ and detuning $\Delta$.  We use a 120~mW diode laser tuned $\Delta \simeq 0.3$~nm above the $^{87}$Rb cooling transition at 780.2~nm, focused to a beam waist of $40~\mu$m.  The 6.8~GHz detuning difference between spin states is sufficient to produce a differential potential up to $U\sub{diff}/h \sim500$~Hz. $\Delta$ was chosen to minimize spontaneous emission but still keep it close enough to resonance to obtain the desired differential potentials without using high intensities, which could exert significant optical forces on the atoms.  We verify this condition by noting that when all of the atoms are in a single spin state, the optical potential produces no noticeable change in the density distribution over the time scale of our experiment.  Because of the elongated aspect ratio of the trap, all of the observable spin dynamics occur only in the axial direction, and the dipole potential can be radially averaged to produce a one-dimensional distribution.  Thus, we need only one axis of spatial control, which we obtain by deflecting the laser with an acousto-optic modulator (AOM) as shown in figure~\ref{fig:apparatus}.

\begin{figure}
\begin{centering}
\leavevmode \epsfxsize=6.3in
\epsffile{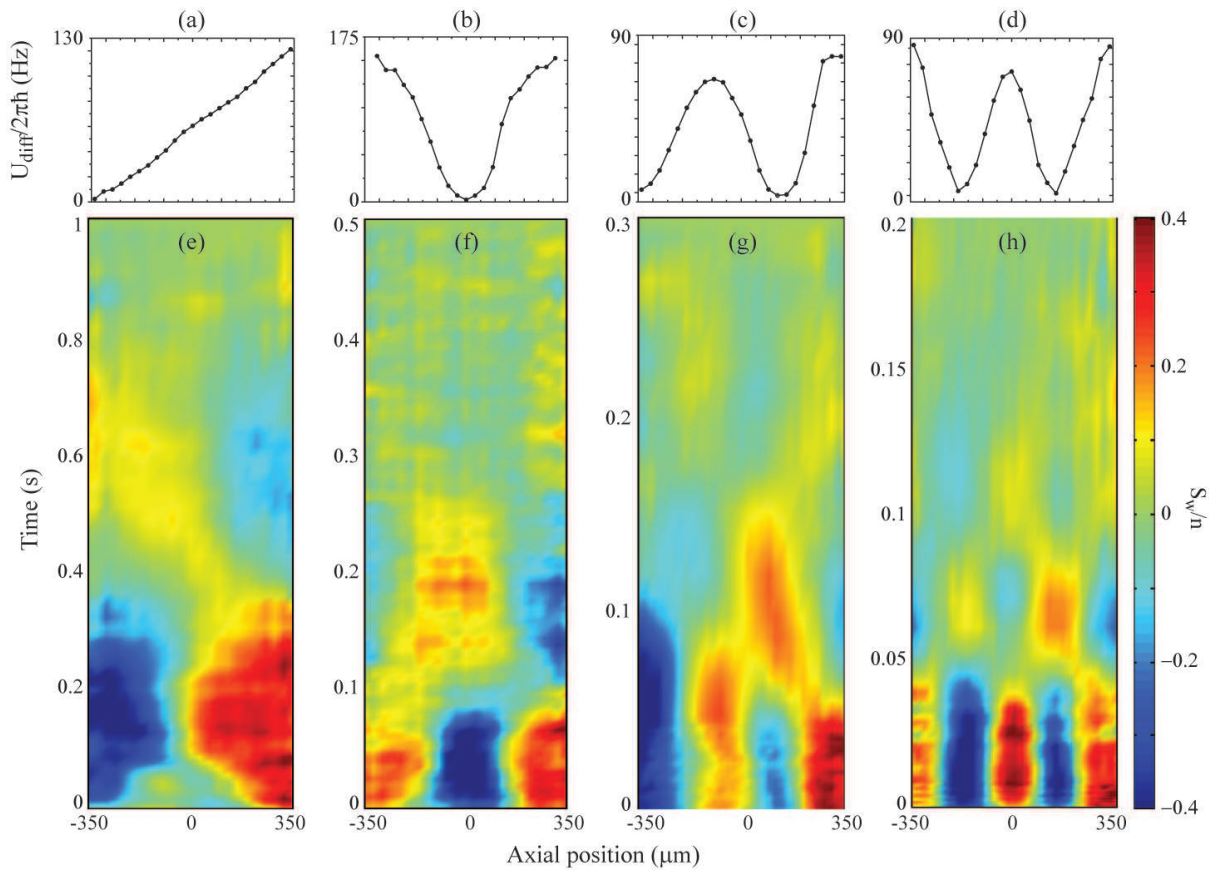}
\caption{\label{fig:modes} $U\sub{diff}/2\pi\hbar$ for the (a) dipole, (b) quadrupole, (c) octupole, and (d) hexadecapole modes.  $U\sub{diff}$ was measured using a $\pi/2-\pi/2$ Ramsey interferometer by varying the $\pi/2$ pulse separation for times up to 5~ms (before atoms will have moved appreciably in the axial direction), measuring the number of atoms returning to the $|1\rangle$ state, dividing the cloud into axial bins for radial averaging, and fitting each bin to a sinusoid to extract the local differential potential energy. Solid lines are to guide the eye.  Each $d$-pole potential has $d$ nodes and is centered and symmetric or antisymmetric.  (e)-(h) show false color images of the spatiotemporal evolution of the normalized longitudinal spin distribution $S_w/n$ for the above multipole modes.  Data is interpolated between bins and time steps.  Red/blue regions represent spatial locations where $S_w$ has rotated out of the transverse plane and represent spin-state segregation.}
\end{centering}
\end{figure}

We tailor spatially varying potentials with arbitrary symmetry by frequency and amplitude modulating (FM and AM) a 120 MHz rf signal applied to the AOM.  The modulation rate of several kHz is significantly faster than all other time scales, including the trap frequencies and collision rates, and this potential can be viewed as a time-averaged, static optical potential.  We create potentials with dipole, quadrupole, octupole, and hexadecapole symmetries ($d=1,2,3,$ and 4 respectively) by adjusting the AM and FM frequencies and phases relative to each other.  For example, to create a dipole differential potential, we use 4~kHz phase-synchronized AM and FM, so that when the dipole beam is swept to one side of the cloud, its amplitude is maximal, while its amplitude is a minimum at the opposite end of the frequency/spatial sweep.  Changing the modulation frequency of the AM and FM relative to each other while also shifting their relative phase produces higher $d$ potentials.  The overall amplitude of the potential is controlled via the laser power.  Figure~\ref{fig:modes}(a)-(d) shows examples of potentials created in this manner.

\section{Optical excitation of spin waves}

To study the intrinsic spin wave modes of the system, we optically imprint these transverse spin profiles (phase profiles) onto the spatially uniform coherent superposition immediately following the $\pi/2$ pulse.  We apply a 5~ms phase imprinting pulse with the same symmetry as the multipole mode we wish to drive, before turning off the optical potential and allowing the spin wave to oscillate freely in an unperturbed uniform potential. After a variable delay time, we measure the 1-D spatial distribution of the longitudinal spin vector $\bi{S} (z)$ by imaging the spatial distribution of the $|1\rangle$ and  $|2\rangle$ states.  We divide the cloud axially into equally sized bins and radially average the atomic distribution in each bin.  Figures~\ref{fig:modes}(e)-(h) show spatiotemporal oscillations of $S_w$ for the first four spin wave modes, plotted as a function of axial position in the cloud and free evolution time.  The longitudinal spin is normalized by the population of each bin, $S_w (z)/n(z)  = \left( N_1 - N_2 \right) / \left(N_2 + N_1 \right)$, to remove number fluctuations and account for trap loss.

\begin{figure}
\begin{centering}
\leavevmode \epsfxsize=3.375in
\epsffile{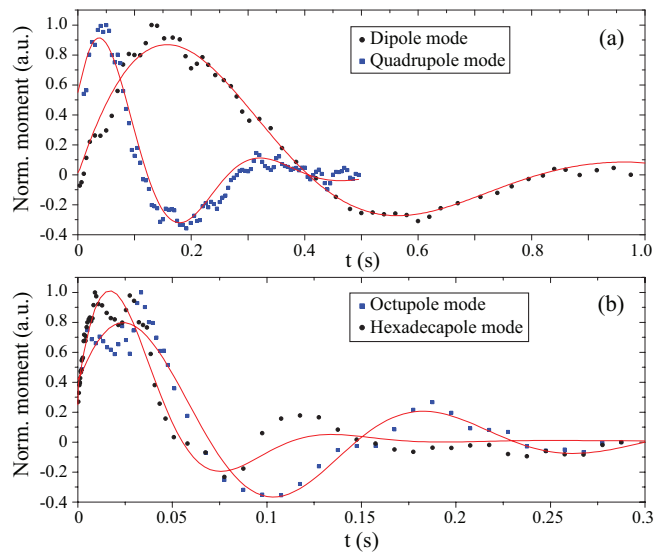}
\caption{\label{fig:moment} Typical normalized spin wave moments, $\langle S_w(z) z^d\rangle_n$, for strongly driven excitations of the (a) dipole ($\fullcircle$) and quadrupole ($\fullsquare$) modes, and (b) octupole ($\fullsquare$) and hexadecapole ($\fullcircle$) modes.  Solid lines are least-squares fits to damped sinusoids, for extracting frequencies and damping rates.}
\end{centering}
\end{figure}

In all cases, the initially transverse spin vectors rapidly develop fluctuations of the longitudinal spin component, which manifests as spin segregation \cite{lewando2002}.  Higher order modes show proportionally higher frequencies due to the symmetry of the differential potentials.  The maximum frequency for any mode $d$ is governed by the trapping frequency according to $f\sub{d} = d\omega\sub{z} / 2\pi$, which occurs in the zero density, collisionless limit.  At higher density, $f\sub{d}$ is still proportional to $d$ but also scales inversely with the exchange collision frequency, which is proportional to density.

The damping rate, $\Gamma\sub{d}$, also increases with mode number $d$, as collisions bring to equilibrium faster those spin distributions that have greater inhomogeneity.  These two differing timescales cancel out somewhat, and the quality factor, $Q\sub{d} = 2\pi f\sub{d}/\Gamma\sub{d}$, varies only slightly for the different modes.  For the $d = 2, 3,$ and 4 modes, $Q\sub{d} \sim 1-4$ and the excitation damps after only one full oscillation.  Because the dipole mode frequency is so low, $f_1 \sim 1$~Hz, it damps via an additional mechanism -- dipolar relaxation of the $|2\rangle$ atoms leading to trap loss and additional shrinking of the spin vector -- and $Q_1$ is smaller.  It is important to note that while we distinguish between the different normal modes of the system based on their symmetry, these modes are not closed and can couple to other spin wave modes.  This coupling is a significant contribution to the damping rate and is a further indicator of the nonlinearity of the system.

\subsection{Amplitude dependence of frequencies}

\begin{figure}
\begin{centering}
\leavevmode \epsfxsize=5.5in
\epsffile{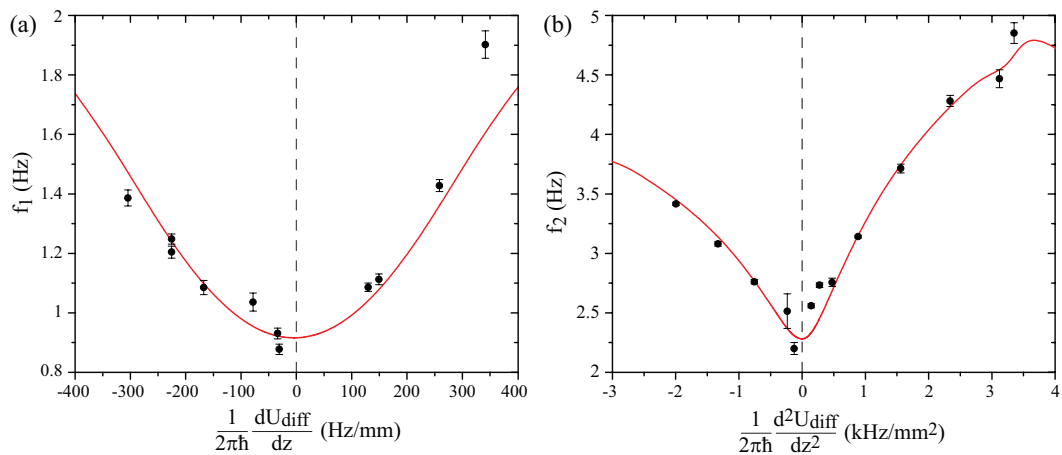}
\caption{\label{fig:frequency} Frequency of (a) dipole and (b) quadrupole spin wave oscillations as a function of excitation amplitude, parametrized by the gradient and curvature of $U\sub{diff}$ at the center of the cloud.  The solid lines are numerical simulations of the quantum Boltzmann equation (see text).  Error bars are statistical and do not include systematic effects such as density calibration and the effects of nonsinusoidal oscillations.}
\end{centering}
\end{figure}

To highlight the nonlinear nature of these large amplitude spin waves, we systematically vary the amplitude of the initial perturbation to the transverse spin by adjusting the amplitude of the 5~ms phase imprinting pulse and studying the resulting spin wave frequencies (figure~\ref{fig:moment}).  The multipole moment of each excitation is calculated according to $\langle S_w(z,t) z^d\rangle_n$ to extract the oscillation frequencies, where the notation $\langle \ldots \rangle_n$ denotes a density weighted average. This density weighting is necessary because the signal-to-noise ratio is lowest on the edges of the cloud where $z^d$ becomes large.  Damped sine waves are then fit to the multipole moments to determine $f_d$ and $\Gamma_d$.  Figure~\ref{fig:frequency} shows the dipole and quadrupole mode frequencies as a function of the amplitude of the initial perturbations.  Negative values in figure~\ref{fig:frequency} are achieved by changing the phase of the FM and AM applied to the phase imprinting AOM.  Other than reversing the sense of the rotation, equation~(\ref{boltzmann}) is unchanged by reversing the sign of $\partial U\sub{diff}/\partial z$ and $\partial^2 U\sub{diff}/\partial z^2$.

For all modes, $f_d$ increases significantly with amplitude, nearly doubling in the case of the quadrupole mode.  As the frequency increases, however, $\Gamma_d$ decreases faster, and $Q_d$ drops for larger perturbations.  The octupole and hexadecapole modes are not studied in detail here, as they are quite sensitive to any imperfections in the optical driving potential.  If the optical potentials do not have the exact symmetry of the desired mode, the $d=3$ or 4 modes quickly damp and degenerate into longer lived $d=1$ or 2 excitations.

To compare these measurements with the spin transport theory, we numerically simulate equation~(\ref{boltzmann}) using an alternating direction implicit finite difference method.  We use the experimentally measured $U\sub{diff}$ (see figure~\ref{fig:modes}(a) and (b) for instance) and account for imaging inefficiencies by adjusting the measured density by $\sim20\%$ to match the low amplitude limit of the experimentally measured frequencies.  For a given $U\sub{diff}$ we calculate the multipole moment from the simulated $\bi{S}(z,t)$ and fit a damped sinusoid to it for comparison with the measured frequencies.  As seen in figure~\ref{fig:frequency}, the simulations show good agreement with the data, indicating that the evolution of the spin vector is well described by the Boltzmann equation.

The primary systematic error in these measurements is loss of atoms from the $|2\rangle$ state due to dipolar relaxation.  As the spin states separate, dipolar relaxation is accelerated, significantly altering the dynamics.  Furthermore, the driving potentials we use are not purely linear or purely quadratic, and lead to excitation amplitudes and patterns that are not completely characterized by a single parameter ($\partial U\sub{diff}/\partial z$ or $\partial^2 U\sub{diff}/\partial z^2$) taken at the center of the trap.  It is important to note that the oscillations of $S_w$ are not truly sinusoidal; however, since $Q_d\sim3$, a sinusoidal fit over a single oscillation is a reasonable characterization of the primary time scale.  We also verify via Monte Carlo simulations that there is no systematic error from fitting damped sinusoids when $f\sub{d}$ and $\Gamma\sub{d}$ are not well separated,  so long as $Q_d > 1$.  However, the fitted frequencies are systematically lowered when gamma becomes larger than than the frequency -- this effect is seen in figure~\ref{fig:frequency}(b) in both the theory and data for the most strongly driven spin waves above 3~kHz/mm$^2$ initial curvature. In this region the fits, while repeatable, are not true measures of an oscillation frequency.

Despite the symmetry of the Boltzmann equation, quadrupole excitations in our experiment are asymmetric with respect to the sign of the curvature of $U\sub{diff}$, as seen in figure~\ref{fig:frequency}(b).  There are several reasons for this asymmetry.  Although we parametrize the potentials by the curvature at the center of the trap, the quadrupole differential potential is not truly quadratic and rolls off near the edges of the cloud (figure~\ref{fig:modes}(b)); this roll off changes shape somewhat when we reverse the sign of $U\sub{diff}$.  This effect is included in our simulations by using the full experimentally determined potentials as inputs, which accounts for some of the positive/negative asymmetry in the frequency.  Another significant effect comes from, again, losses of the $|2\rangle$ state via dipolar relaxation collisions.  When $|2\rangle$ segregates to the trap center (positive curvature), dipolar loss slightly speeds up the return of the $|1\rangle$ state to the center of the trap.  However, for negative curvature, $|2\rangle$ segregates to the outside of the trap and preferentially leaves the trap, making the return oscillation appear slower since the spin current is not conserved.  We account for this effect at each time step of the simulation by adding a small phenomenological loss to the projection of $\bi{S}$ onto $|2\rangle$ proportional to the local $|2\rangle$ state density.  With these modifications, we obtain good agreement between the simulations and measurements, both showing strong nonlinear behavior.  The linear regime, where the frequency curve flattens, occurs at extremely small excitation amplitude and is almost nonexistent on this scale.  We are not able to drive small enough amplitude spin waves to explore this limit.

\subsection{Short time behavior}

The behavior of the spin vector shortly after the optical excitation commences also highlights the nonlinear nature of these spin waves.  Both the small amplitude linearized moment method of \cite{nikuni2003} and the near equilibrium hydrodynamic approach of \cite{levitov2002} predict that $S_w$ initially grows as $t^2$.  However, equation~(\ref{boltzmann}) can be expanded for $t <\tau\sub{col}$, the mean collision time \cite{williams2002,fuchs2003}.  In this limit, the initial growth of $S_w$ goes as $t^4$:
\begin{equation} \label{smallt}
\frac{S_w}{n} = \frac{gn}{48\hbar} \frac{k\sub{B} T}{m} \left[ \frac{d^2U\sub{diff}}{dz^2} - \frac{2z}{\sigma_z^2} \frac{dU\sub{diff}}{dz} \right] t^4,
\end{equation}
where $\sigma_z$ is the axial Gaussian width of the atomic distribution.

\begin{figure}
\begin{centering}
\leavevmode \epsfxsize=3.375in
\epsffile{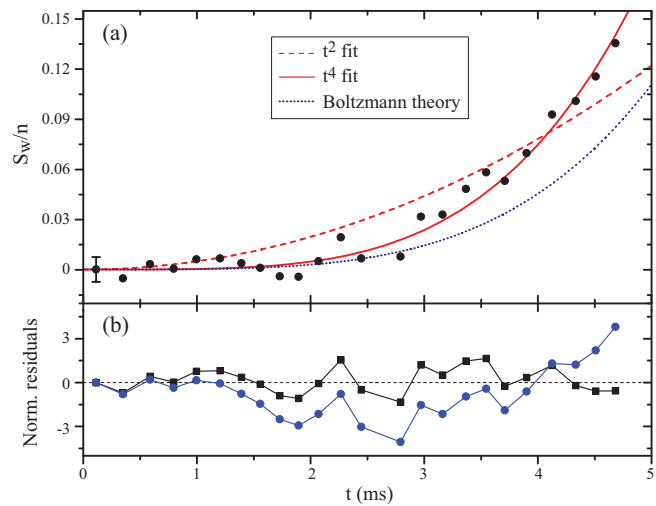}
\caption{\label{fig:smalltime} (a) Nonlinear behavior in the initial growth of $S_w$ (normalized) at the center of the cloud.  Nonlinear least-squares fits to quartic (solid line) and quadratic (dashed line) are shown, along with the prediction of equation~(\ref{smallt}), the Boltzmann transport equation (dotted line).  All points are an average of three experimental cycles, and the first point shows typical statistical error bars.  (b) Normalized residuals from the quadratic ($\fullcircle$) and quartic fits ($\fullsquare$) with solid lines to guide the eye.  The residuals show a clear systematic trend for the quadratic fit (reduced $\chi^2 = 3.8$), while the quartic fit residuals ($\chi^2 = 1.3$) indicate that a quartic increase in $S_w$ occurs.}
\end{centering}
\end{figure}

Figure~\ref{fig:smalltime} shows $S_w$ in the center of the cloud for a hexadecapole excitation similar to that shown in figures~\ref{fig:modes}(d) and (h) for times shorter than $\tau\sub{col} \simeq 12$~ms.  For this study we use a strong optical excitation to maximize nonlinearities and measure $S_w$ at a short time $t$ after the start of the optical excitation pulse, all of which occurs before free evolution time shown in figure~\ref{fig:modes}.  We bin pixels in the center 23~$\mu$m of the cloud image for averaging and normalize $S_w$ by the number of atoms contained within that bin.  We fit both $t^2$ and $t^4$ power laws to $S_w$.  Although the $\sim5\%$ shot-to-shot number fluctuations contribute significant noise to a small $S_w$, the initial spin growth clearly does not grow quadratically and is better described by a quartic rise.

The interpretation for this behavior is that the superposition must first build up phase gradients before coherent spin currents can develop, and only then does the ISRE begins to convert transverse phase gradients into longitudinal spin rotations.  This multi-step process takes longer to initiate but rapidly accelerates once sufficient phase gradient has accumulated in what \cite{fuchs2003} likened to a ``density shock'' rather than hydrodynamic or collisionless flow.  The observed initial growth rate is even faster than that predicted by equation~(\ref{smallt}), which is most likely explained by the fact that the spin vectors continue to evolve during the several hundreds of microseconds imaging sequence prior to the image acquisition itself, as well as possible small errors in the calibration of density and $U\sub{diff}$ used in the simulations.  As spin vectors rotate significantly out of the transverse plane and $S_w$ gets large, the rate of increase slows down and becomes quadratic, approaching the sinusoidal behavior seen in figure~\ref{fig:moment}.

\section{Conclusion}

We have demonstrated a versatile technique for exciting arbitrary modes of spin waves with arbitrary amplitudes.  We have used this optical excitation technique to highlight the nonlinear behavior of an ultracold trapped atomic gas.  Because the typical operating regime for spin waves in this system tends to be squarely between the hydrodynamic and collisionless limits, the behavior of spin waves is more complicated than that described by linear or small amplitude approximations. In this case, it is almost impossible to excite purely linear spin waves, as even modest perturbations drive the sensitive atomic system into large amplitude fluctuations with significant nonlinear effects, including amplitude and time dependent changes in excitation frequency, increased damping, coupling between modes, and deviations from pure sinusoidal behavior.  Furthermore the nature of the ISRE is complicated enough that even in the collisionless limit, the momentum cannot always be integrated out of the Boltztmann equation, and the dynamics are still not simple \cite{thomas2008,deutsch2010,laloe2009}.  In either case, the ISRE and spin waves continue to offer rich and interesting examples of out-of-equilibrium quantum systems.

\section{Acknowledgments}

This work was supported by NSERC and the CFI.  L.Z. acknowledges support from the SFU-VPR Fellowship Program.

\end{document}